\documentclass[sigconf,screen,nonacm]{acmart}

\AtBeginDocument{%
  }

\definecolor{lightgray}{rgb}{.9,.9,.9}
\newcommand\thesis[1]{\noindent \begin{center} \vspace{-.4em} \hspace*{0.5ex} \colorbox{lightgray}{\parbox{0.95\columnwidth}{#1}} \vspace{-.7em} \hspace*{0.5ex} \end{center}}

\usepackage{xcolor}
\usepackage{tabularx}
\usepackage{makecell}
\usepackage{paralist} 
	
\usepackage{soul}

\begin{document}
\newcommand\numthesis{11 }

\title{Tapping into the Natural Language System with Artificial Languages when Learning Programming}

\author{Elisa Madeleine Hartmann}
\affiliation{%
  \institution{Chemnitz University of Technology}
  \streetaddress{Straße der Nationen 62}
  \city{Chemnitz}
   \state{Saxony}
  \country{Germany}}

\author{Annabelle Bergum}
\affiliation{%
 \institution{Graduate School of Computer Science,\\Saarland University, Saarland Informatics Campus}
 \city{Saarbr\"ucken}
 \country{Germany}}

\author{Dominik Gorgosch}
\affiliation{%
 \institution{Chemnitz University of Technology}
 \streetaddress{Straße der Nationen 62}
 \city{Chemnitz}
 \state{Saxony}
 \country{Germany}}

\author{Norman Peitek}
\affiliation{%
 \institution{Saarland University, Saarland Informatics Campus}
 \streetaddress{}
 \city{Saarbr\"ucken}
 \country{Germany}}

\author{Sven Apel}
\affiliation{%
 \institution{Saarland University, Saarland Informatics Campus}
 \city{Saarbr\"ucken}
 \country{Germany}}

\author{Janet Siegmund}
\affiliation{%
 \institution{Chemnitz University of Technology}
 \streetaddress{Straße der Nationen 62}
 \city{Chemnitz}
 \state{Saxony}
 \country{Germany}}


\begin{abstract}

\textit{Background}: In times when the ability to program is becoming increasingly important, it is still difficult to teach students to become successful programmers. One remarkable aspect are recent findings from neuro-imaging studies, which suggest a consistent role of language competency of novice programmers when they learn programming. Thus, for effectively teaching programming, it might be beneficial to draw from linguistic research, especially from foreign language acquisition. 
\\
\textit{Objective}: The goal of this study is to investigate the feasibility of this idea, such that we can enhance learning programming by activating language learning mechanisms. \\
\textit{Method}: To this end, we conducted an empirical study, in which we taught one group of students an artificial language, while another group received an introduction into Git as control condition, before we taught both groups basic programming knowledge in a programming course.\\
\textit{Result}: We observed that the training of the artificial language can be easily integrated into our curriculum. Furthermore, we observed that language learning strategies were activated and that participants perceived similarities between learning the artificial language and the programming language. However, within the context of our study, we did not find a significant benefit for programming competency when students learned an artificial language first.\\
\textit{Conclusion}: Our study lays the methodological foundation to explore the use of natural language acquisition research and expand this field step by step. We report our experience here to guide research and to open up the possibilities from the field of linguistic research to improve programming acquisition.

\end{abstract}


\maketitle

\section{Introduction}

Learning to program is difficult \cite{luxton-reilly_learning_2016}. At universities, students have been failing their programming courses for decades~\cite{bennedsen_failure_2007, watson_failure_2014, bennedsen_failure_2019, simon_pass_2019}, and even for those who pass their courses, it is unclear what programming skills they have actually developed~\cite{soloway_tapping_1982, mccracken_multi-national_2001, lister_multi-national_2004, lister_asssesment_2006, utting_fresh_2013}. 
The problem is that, despite decades of research and dedicated research communities, we still do not know how to effectively teach programming~\cite{bennedsen_failure_2007,bennedsen_failure_2019,soloway_tapping_1982, mccracken_multi-national_2001, lister_multi-national_2004, utting_fresh_2013}.

In recent neuro-imaging studies on programming, researchers have identified consistent activation of brain areas related to language processing~\cite{siegmund2014understanding, prat2020relating, Hongo22, ICSE21, krueger2020neurological}, indicating a closeness of programming and language processing. This appears reasonable, as there are similarities between the process of learning a programming language and the process of learning a natural language: In both cases, learners need to develop an understanding of the rules and structure of a language, as well as the ability to use it to communicate and accomplish specific tasks. Few studies already observed a similarity, for example, in the acquisition of syntax and semantics~\cite{tshukudu2020understanding,shrestha2020}. Thus, it seems opportune to employ a language learning lens to programming. This way, we can tap into the widely explored field of natural language learning. 

A typical way to study language learning uses artificial grammars and languages~\cite{kinder2000learning, knowlton1996artificial, friederici2002brain, opitz2003interactions, opitz2011timing, reber1967implicit, ettlinger2016relationship, reber1989implicit}. The advantage compared to using natural languages is that, in an artificial language, the impact of an already learned language can be reduced.
In addition, artificial languages are much smaller than natural languages, so they can be learned in a short time (e.g., within a few hours). Neuro-imaging studies based on artificial languages show an involvement in activation of typical language processing areas~\cite{petersson2012artificial}, demonstrating that artificial languages are a valid substitute for studying natural languages.

We take advantage of this connection and draw on this experience for a better understanding and training of programming. Our first goal is to evaluate whether the approach of building on artificial grammars and languages can be transferred to programming experiments. Specifically, we evaluate whether we can train language learning aspects with an artificial language before students start learning programming, such that learning programming becomes easier. Our goal is to activate language learning strategies by learning the artificial language. As the chosen language consists of syntactical constructs, apart from semantic activation, we investigate the transfer from the language learning strategies to syntactical constructs as i.e. if-conditions. Therefore, our focus does not yet lie on programming competencies, but on the learning process itself, which should be simplified. With the results of this study we will be able to make additional assumptions on the learning process and its link to program comprehension.

We expect that training strategies used to learn artificial languages taken from the repertoire of language learning strategies have a positive effect on learning programming. These strategies include the transfer of pattern templates of already known structures to a new language~\cite{selinker1969language, schachter1983new}. An example is \emph{inversion} in German, which describes a change in the word order of a main clause when, for example, a temporal adverb is inserted ahead: In a simple main sentence, the verb follows the noun, for example: ''Die Katze \emph{(noun)} ist \emph{(verb)} schwarz \emph{(adjective)}.'' (``The cat \emph{(noun)} is \emph{(verb)} black \emph{(adjective)}.''). With an added temporal adverb, the noun and verb swap positions: ''Heute \emph{(adverb)} ist \emph{(verb)} die Katze \emph{(noun)} schwarz \emph{(adjective)}.'' (``Today \emph{(adverb)} is \emph{(verb)} the cat \emph{(noun)} black \emph{(adjective)}.''). That a transfer takes place can be observed in the fact that
this specialty in syntax often causes problems for English-speaking learners of German, who form the sentence according to English syntax without active verb displacement by adding the temporal adverb without changing anything in the rest of the syntax (e.g., ``The cat is black'' vs.\ ``Today, the cat is black'')~\cite{pienemann1988constructing}. These so-called \emph{interferences} illustrate that knowledge of a native or already acquired language transfer to a (new) foreign language (and vice versa).
It also demonstrates that consistent or similar structures can be acquired with less effort. This has also been observed when programmers learn a new programming language, such that natural language transfer concepts also come into play when learning a new programming language and learning to program, including observable interference~\cite{tshukudu2020understanding, shrestha2020}.

This might also apply when learning programming is preceded by learning an artificial language, such that areas in the brain related to language processing might adapt to processing a programming language. This adaptability is an inherent property of the brain to adapt to new tasks: For example, the fusiform face area is involved in face recognition, but also has adapted in bird experts to recognize birds, and in car experts to recognize cars (but not vice versa)~\cite{Gauthier00}. Just like the fusiform face area can adapt to other tasks, the language processing area(s) are active during programming. Thus, tapping into the language system might help in learning programming.

To evaluate the feasibility of teaching an artificial language as a vehicle for transfer to train beginner programmers, we explore the experiment design landscape. As a first step, we developed a programming course for beginner programmers that starts with learning the artificial language \emph{Brocanto} before learning programming. \emph{Brocanto} is used in linguistic neuroscience studies, and has proved successful in natural language learning studies~\cite{opitz2003interactions, batterink2013implicit, friederici2002brain, opitz2004brain, opitz2011timing, cooper2019influence}. It is structurally adapted to natural languages and consists of nominal phrases connected by verb phrases, both of which are constructed with pseudowords. Notably, \emph{Brocanto} is also similar to programming languages, in that it has a restricted syntax and a fixed set of keywords. To evaluate the effect of learning \emph{Brocanto} on learning programming, we conducted a study, in which one group learned \emph{Brocanto} before programming (\emph{Brocanto} group), and a control group received an introduction to Git before programming (Git group). All students from the \emph{Brocanto} group successfully learned it within one hour. Additionally, we asked the students after the course whether they recall using specific strategies in learning the artificial language, and students mentioned that they oriented themselves along certain aspects of the language (from single words to phrase structure forming markers).

Despite hints that language learning strategies were activated in the \emph{Brocanto} group, there was no statistically significant difference regarding programming learning within the context of our study. However, our study spanned only one week, and an effect might be observable only at a later point. For further assessment, we take a look at the results of programming tasks from the first semester course \emph{Algorithms and Programming} after the first semester to get an extended impression.
Nevertheless, our experiment design provides a framework for conducting further programming learning studies that are facilitated by language learning, and the statements of the participants motivate us to further explore this experimental landscape. Based on the results, we describe a road map to continue this line of work by formulating \numthesis conjectures that are promising for future research. 

In summary, we make the following contributions:

\setdefaultleftmargin{1em}{}{}{}{}{}
\begin{compactitem}
    \item An experiment design using an artificial language to improve learning programming
    \item \numthesis conjectures to guide future research
    \item A replication package containing all information to replicate and adapt our study
\end{compactitem}

\section{Brocanto}
\label{sec:background}

In this section, we provide an overview of \emph{Brocanto}, its use cases, and how we used it in our study. We start by discussing why it is a suitable artificial language in the context of programming learning.

\subsection{Motivation}
In experiments on natural languages, artificial languages have proven to be valid substitutes for natural languages~\cite{opitz2003interactions, opitz2004brain}. 
As we put the natural language (learning) into focus in this study, we decided against a treatment with another programming language. Therefore, it would have been a study on transfer.
For the same reason, we also did not choose to contradict syntax exercises with Python, so called 'Drill tasks' as we wanted to implicitly activate learning mechanisms that are used in natural language learning.
Our goal is to evaluate whether learning an artificial language can act as a transfer medium of concepts from a natural language to a programming language. This way, we tap into the language system to make programming learning easier. It is reasonable to believe that this is possible, because artificial grammars, much like programming languages, have strict structure rules and are typically fast to learn (i.e., within a few hours)~\cite{kinder2000learning, knowlton1996artificial,opitz2003interactions}.

\emph{Brocanto} consists of universal principles of natural languages, such as (pseudo)words, phrases, and syntax rules, which also count for programming languages. Thus, \emph{Brocanto} can act as a bridge element between natural languages and programming languages. Additionally, \emph{Brocanto} supports transfer, so that knowledge from a known natural language is applied in learning \emph{Brocanto}~\cite{Thorndike01}. This can be seen in phrase structure rules and finite-state grammar, which are language learning methods that are active in the acquisition of an artificial language~\cite{reber1967implicit, reber1989implicit}.

First, the phrase structure rules method describes that language learning entails breaking down a language's constituent parts, also known as syntactic categories, into its phrases. Since \emph{Brocanto} has a phrase structure grammar, a transfer of this knowledge of natural languages can be applied to learning \emph{Brocanto}. In other words, when learning \emph{Brocanto}, students decompose \emph{Brocanto} sentences in substructures as part of the learning process, which also happens during processing natural languages~\cite{reber1967implicit, reber1989implicit}.

Second, the finite state grammar focuses on the probability of transition of individual elements, which increases with an increasing similarity between languages. For example, Brooks and Vokey found that, with similarity-based learning, students apply grammatical knowledge to a new set of letters (from artificial grammars)~\cite{brooks1991abstract}. Thus, what is similar is also more easily transferred between languages. Both learning strategies have been observed in studies with \emph{Brocanto}, making it particularly suitable for our experiment.

\subsection{Definition}
\label{sec:defBrocanto}

\begin{figure*}[htbp]
    \centering
    \includegraphics[width = \textwidth]{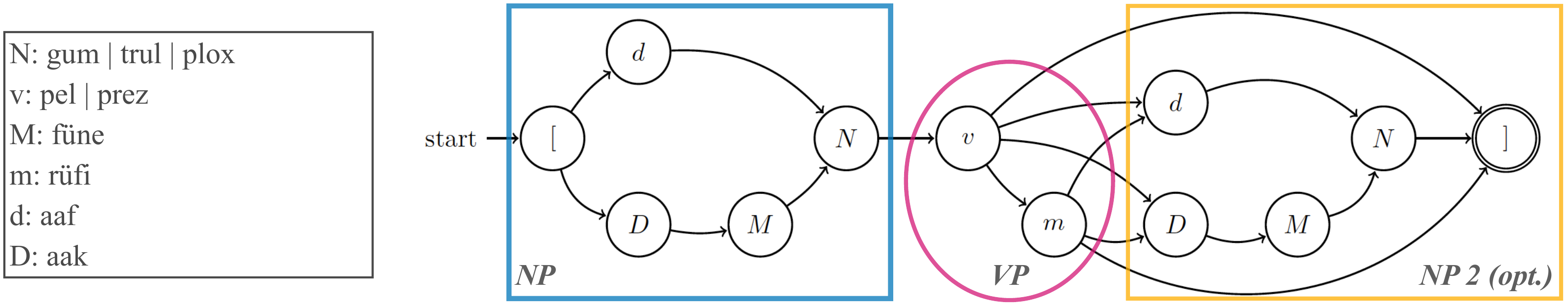}
    \caption{\emph{Brocanto} Transition Diagram / Automat}
    \label{fig}
\label{fig:Brocanto}
\end{figure*}

\begin{table*}[t]
\caption{Violation Overview}
\centering
 \begin{tabularx}{\textwidth}{p{5cm}|r|p{4cm}|r}
 \toprule
    Violation & Structure Example & Brocanto Example  &  Natural Language Example \\
\midrule
No Violation &   NP +VP +NP &            aak füne trul prez aaf plox  &            My black cat makes the noise.  \\
Phrase order violations &  \textbf{NP} + \textbf{NP} + VP &  aak füne trul \textbf{aaf plox}  prez & My black cat \textbf{the noise} makes.  \\
Determiner-noun agreement violations &   \textbf{d}+\textbf{M}+N+v+d+N  &        \textbf{aaf füne} trul prez aaf plox &            \textbf{ My the} black cat makes the noise. \\
Word category repetitions &  D+\textbf{N}+M+\textbf{N}+v+d+N     &          aak \textbf{plox} füne trul prez aaf plox &             My \textbf{noise} black \textbf{cat} makes the noise. \\
\bottomrule
\end{tabularx}
\label{tab_brocanto_overview}
\end{table*}

With its different nominal and verb phrase formations, \emph{Brocanto} allows the definition of main clauses with a subject-verb-[object] structure. Formally, \emph{Brocanto} is defined as:

\begin{itemize}
    \item Start symbol S
    \item S := NP $\circ$  VP $|$ NP $\circ$  VP $\circ$  NP    \hfill $\vartriangleright$ \textit{Sentence}
    \item NP := d $\circ$  N $|$ D $\circ$  M $\circ$  N     \hfill $\vartriangleright$ \textit{Nominal Phrase}
    \item VP := v $|$ v $\circ$  m          \hfill $\vartriangleright$ \textit{Verb Phrase}
    \item N := \textsf{gum} $|$ \textsf{trul} $|$ \textsf{plox} $|$ \textsf{tok}     \hfill $\vartriangleright$ \textit{Noun}
    \item v := \textsf{pel} $|$ \textsf{prez} $|$ \textsf{glif} $|$ \textsf{rix}      \hfill $\vartriangleright$\textit{Verb}
    \item M := \textsf{böke} $|$ \textsf{füne}         \hfill $\vartriangleright$\textit{Adjective}
    \item m := \textsf{nöri} $|$ \textsf{rüfi}          \hfill $\vartriangleright$\textit{Adverb, Verb-Suffix (Conjugation)}
    \item d := \textsf{aaf}          \hfill $\vartriangleright$\textit{Definite Article, Pronoun}
    \item D := \textsf{aak}          \hfill $\vartriangleright$\textit{Definite Article, Pronoun}
\end{itemize}

The nominal phrase (\textsf{NP}) consists of a determiner (\textsf{D, d}), optionally an adjective (\textsf{M}), and a noun (\textsf{N}). Thus, a nominal phrase can be written formally as (\textsf{dN} or \textsf{DMN}). Verb phrases (\textsf{VP}) consist of a verb (\textsf{v}), and optionally an adverb or verb-suffix (\textsf{m}), and can be written as \textsf{v} or \textsf{vm}. Thus, the sentence \textsf{aaf~plox~prez~aak~böke~trul} is correct according to \emph{Brocanto} syntax, but \textsf{aaf~plox~prez~böke~trul} is not, because \textsf{prez~böke~trul} does not follow the construction of nominal phrases, as \textsf{böke trul} does not have a determiner (\textsf{aaf} or \textsf{aak}).

The standard protocol to teach artificial languages is to confront participants with grammatically correct and incorrect sentences~\cite{reber1967implicit, reber1989implicit,kinder2000learning}. There are three different types of grammatical violations that have proved successful for learning \emph{Brocanto}~\cite{opitz2004brain, opitz2011timing}: Phrase order violations, determiner-noun agreement violations, and word category repetitions. 

\paragraph{Phrase order violation} The valid phrase order of a sentence is violated when the order of phrases does not conform to \textsf{NP+VP} or \textsf{NP+VP+NP}. For example, if the verb phrase is put at the beginning of a sentence, the phrase order is violated~(see Table~\ref{tab_brocanto_overview} for an example).

\paragraph{Determiner-noun agreement violation} \emph{Brocanto} distinguishes between two determiners: one that can form the nominal phrase with the noun alone (\textsf{d}) and another that can form the nominal phrase only with the addition of an adjective (\textsf{D+M}). The determiner \textsf{D} must be followed by an adjective \textsf{M} before the noun \textsf{N}. The determiner \textsf{d} can only be followed by a noun \textsf{N}. Ignoring this rule and adding an adjective \textsf{M} to the determiner \textsf{d} leads to an agreement violation~(cf.\ Table~\ref{tab_brocanto_overview}).

\paragraph{Word category repetition} Since \emph{Brocanto} has fixed word categories, each word can be assigned a concrete position within a phrase; within phrases, word categories cannot repeat. If two nouns occur in one nominal phrase, the structure of that phrase is violated; it does not matter at which position the word category repetition is placed (cf.\ Table~\ref{tab_brocanto_overview}). 

We use these three kinds of violations in our experiment.

\subsection{Adaptation}

For our experiment, we adapted \emph{Brocanto} to fit within the time frame of our study, as the full version takes several hours to learn. The stimulus material that we created followed the treatment according to Opitz 2011~\cite{opitz2011timing} included all pseudowords from \emph{Brocanto}. The pilot showed that in the context of our short treatment, using sentences with all possible 14 pseudowords leads to cognitive overload. So we adapted the material to Opitz \cite{opitz2003interactions}, in which the procedure more closely matched our time period.
Specifically, we restricted the sentence length to 5 to 8 words and shortened the set of \emph{Brocanto}'s pseudowords. From the 14 pseudowords, 
we included 9 in our study (3x\textsf{N}, 2x\textsf{v}, 1x\textsf{M}, 1x\textsf{m}, 1x\textsf{d}, 1x\textsf{D}; cf.\ Fig.\ \ref{fig:Brocanto}).


\section{Experiment Design}
Having introduced \emph{Brocanto}, we present the design of our study. Details of questionnaires, course material, and data are available at the project's Web site\footnote{\url{https://github.com/brains-on-code/Tapping-into-the-Natural-Language-System-Using-Artificial-Languages-when-Learning-Programming}}.

\subsection{Research Question}

To guide our experiment design, we pose the following research questions:

\begin{itemize}

\item[\textbf{RQ\_1}] Can we feasibly integrate artificial language learning into our curriculum?

\item[\textbf{RQ\_2}] Does learning the artificial language \emph{Brocanto} benefit learning how to program in a beginner programming course compared to learning Git?

\end{itemize}

To measure a possible effect of learning \emph{Brocanto} on learning programming, we follow a between-subjects design. The independent variable is the treatment each group receives: One group learns \emph{Brocanto}, the other group gets an introduction to Git as control, so that not only the extra time students spend with learning something (new) affects programming learning.

To operationalize programming skill, we conduct programming tests at the beginning (pretest) and end (posttest) of the course and measure the number of correct answers in each test. To control for programming experience and knowledge, we divide the participants into two comparable groups based on a questionnaire applied before the study. We explain the details of the tests and questionnaire next. 






\subsection{Task and Material}

This section explains the questionnaires, tests, and material in the order in which they appeared in our study, which was run over one week.

\paragraph{Pretest}
The first day of the study included an introduction round to welcome participants and to assess previous programming experience. Then, students completed an adapted version of an established programming experience questionnaire~\cite{siegmund2014measuring} and a test to assess existing programming skills~\cite{ahadi2014falling}. The pretest by Ahadi and others provided us with an extended assessment of students' experience by using an applied variable swap to evaluate whether basic programming concepts are already known. It consists of 8 tasks, each of which can either be solved correctly (1 point), or incorrectly (0 points), leading to a maximum of 8 points in this test. The results from the questionnaire and pretest were the basis to create two comparable groups regarding programming skill. On the next day, the two groups individually received their treatment in \emph{Brocanto} or an introduction to Git.

\paragraph{Treatment Brocanto}

\begin{figure*}[htbp]
    \centering
    \includegraphics[width = .95\textwidth]{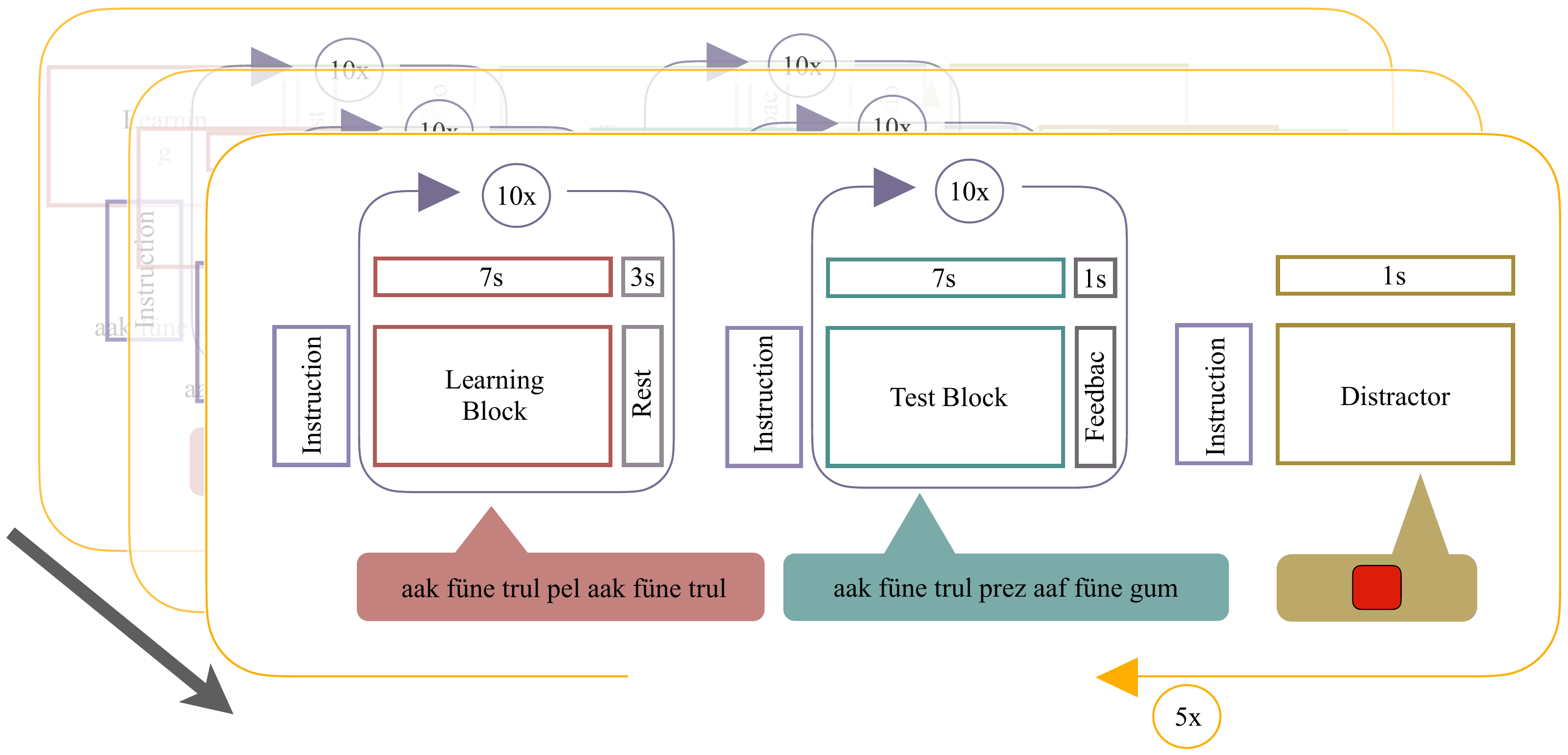}
    \caption{Illustration of one experiment cycle}
    \label{fig:Brocanto Cycle}
\end{figure*}

\begin{figure*}[t]
    \centering
    \includegraphics[width = \textwidth]{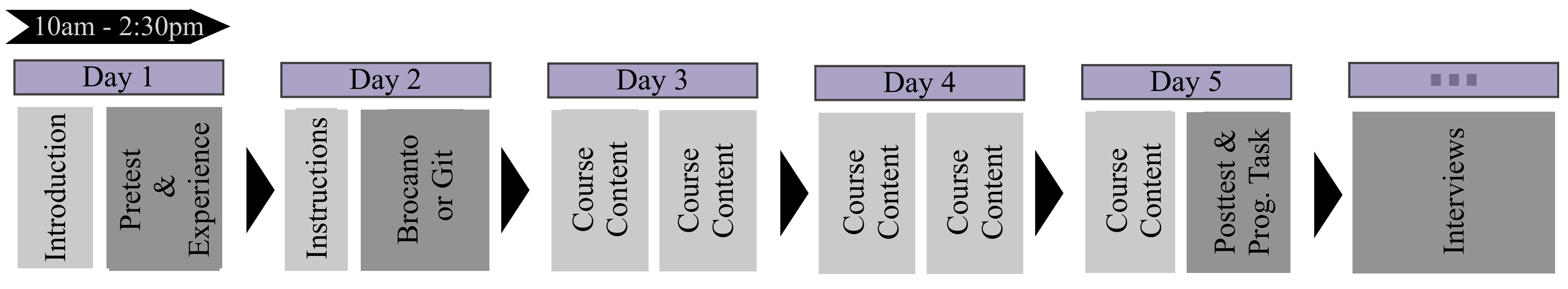}
    \caption{Procedure of Programming Course}
    \label{fig:Programming Course}
\end{figure*}

For the treatment, we created 60 correct sentences and 60 incorrect sentences, 20 of each of the three types of violation (cf.\ Section~\ref{sec:defBrocanto}). For teaching \emph{Brocanto}, we followed standard procedure with three types of blocks that were repeated: Learning, testing, and distractor (see Figure~\ref{fig:Brocanto} for an overview)~\cite{opitz2004brain,knowlton1996artificial,kinder2000learning}. In the \emph{learning block}, participants see 10 correct sentences, each for 7 seconds, and are instructed to deduct the underlying grammar rules. Between each of the 10 sentences, a fixation cross was shown for 2 seconds in the center of the screen.

In the \emph{test block}, participants see 10 sentences, each for 7 seconds. 5 of the sentences are correct, and 5 incorrect, containing any of the three violations, (cf.~Section~\ref{sec:background}), in random order. Of the correct sentences, around half were already shown in the learning block, and the other half were entirely new. The sentences with violations were new to the participants, and no violated versions of the sentences shown in the learning block. Due to \emph{Brocanto's} limited repertoire of words and phrases, sentences that were shown in correct form in the \emph{learning block} may appear in violated form in one of the \emph{test blocks}. Because of the violations, the sentences are not comparable. Participants were instructed to intuitively decide as fast as possible whether the sentence was correct or incorrect by pressing the right (correct) or left (incorrect) arrow key. After each decision, participants immediately received feedback for 1 second. If participants did not respond within the 7 seconds, the decision was logged as incorrect and participants could proceed by pressing the space bar.

The purpose of the \emph{distractor block} was to inhibit memorization of grammar rules and correct/incorrect sentences, so participants still deduct them in subsequent blocks. To this end, the \emph{distractor block} contained a forced key choice, in which participants saw one of two colored geometrical shapes. For a red square, participants should press \emph{y}, followed by \emph{space}, and for a yellow circle, \emph{n}, followed by \emph{space}. 

These three blocks were repeated five times, summarized into one cycle. The experiment consisted of 3 cycles. After each cycle, participants could take a short break (e.g., loosen their hands) and continue by pressing the space bar. The two geometrical shapes that participants had to react to in the \emph{distractor block} did not change. Since it is easy to remember the two key choices, the \emph{distractor block} might no longer distract properly towards the end of the experiment. For this reason, the instruction on how to react was omitted in the last 3 cycles. The participants had to react from memory with the correct key choice when seeing the respective geometric shape. They were not informed beforehand that the instruction would be omitted at the end. The active confusion was to support that they had to fall back on implicit learning that was trained in the experiment in the last \emph{test blocks}.

Participants completed a short warm-up cycle of the same setting consisting of 4 combinations of an artificial grammar of letters to familiarize themselves with the experiment setting.  

\paragraph{Control Git} 
Participants received a brief introduction to version control systems and the advantages of Git over clouds or local version control systems. Afterward, we taught the basic concepts and commands of Git (e.g.,  push, pull, commit). We showed them how to create new projects, add content, and create new branches with GitHub Desktop. During the creation of the new branches, we explained potential merge conflicts. After this introduction, the participants were able to create a new project, clone a project, and use the standard Git-commands. The participants did not practice programming commands in the process. 

\paragraph{Programming Instructions}

The remaining days, both groups met again in the same room and received programming training in Python, covering basic aspects, such as variables, data types, and control structures, adapted from a course of Xie and others~\cite{xie_theory_2019}. This included intermediate programming assignments to asses whether the concepts were understood.

\setlength{\tabcolsep}{3pt}
\begin{table*}[t]
\caption{Pretest, exercises per programming construct and the posttest, as well as the posttest programming task and its percentage correctness divided by test and control group. The programming posttest is presented on a scale between 0 and 2.}
\centering
\begin{tabularx}{\textwidth}{@{}lr|rrrrrrr|rrr@{}}
\toprule
& & \multicolumn{7}{c|}{Exercise Scores} & \multicolumn{3}{c}{Posttests Scores} \\
\makecell[l]{Parti-\\cipant} & Pretest & Data types &  Variables  & Arithmetic  &  Print  &  Logical  & Conditionals &  Loops & Overall & \makecell[l]{Program-\\ming} & AaP\\
\midrule
PG1 &   88 &            100 &             88 &  86 &         71 &                95 &   60 &         81 &  100 & 2 & -\\
PG2 &  63 &  100 &            100 &             100 &        100 &                85 &                   100 &         89 & 100 & 2 & 42 \\
PG3 &   0   &           100 &             96 &              81 &         71 &                90 & 68 & 66 & 69 & 0 & - \\
PG4 &  63     &          100 &             92 &              89 &         57 &                90 &                    92 &         57 & 88 & 2 & 41\\
PG5 & 100 & 100 & 100 & 100 & 100 & 100 & 100 & 15\textsuperscript{*} & 100 & 2 & 11 \\
\midrule
Mean & 63 & 100 & 95  & 91 & 80 & 92 & 84 & 73 & 91 & 1.6 & 31.3\\
\midrule
PB1 &  31 & 100 & 96 &  100 & 100 &  85 & 100 &  81 & 100 & 2 & -\\
PB2 &            38 & 100 & 96 & 94 & 100 & 85 & 96 & 78 &88 & 2 & -\\
PB3 &             0 &                 100 &                 100 &                   94 &              71 &                80 &                   100 &              88 &            100 & 2 & - \\
PB4 &            75 &                 100 &                 100 &                   97 &             100 &                95 &                   100 &              98 &             75 & 2 & - \\
PB5 &            94 &                 100 &                 100 &                  100 &             100 &                95 &                    96 &             100 &             81 & 2 & - \\
PB6 &            94 &                 100 &                  96 &                   97 &             100 &                95 &                   100 &              99 &            100 & 2 & 28 \\
PB7 &            25 &                 100 &                  96 &                  100 &              71 &                85 &                   100 &              63 &             75 & 1 & - \\
PB8 & 100 & 92 & 88 & -\textsuperscript{+} & -\textsuperscript{+} & 100 & 96 & 85 & 100 & 2 & 34 \\
\midrule
Mean & 57 & 99 & 97 & 97 & 92 & 90 & 99 & 87 & 90 & 1.9 & 31 \\
\bottomrule
\multicolumn{6}{l}{\footnotesize * Did not solve all tasks regarding loops due to illness and is not included in this mean.} & 
\multicolumn{3}{c}{\footnotesize + Did not attend these assignments.} & 
\multicolumn{3}{c}{\footnotesize - Cancelled the semester course.}\\
\end{tabularx}
\label{tab_mean_std}
\end{table*}

\paragraph{Posttest}
On the last day of the programming course, a posttest was administered. It consisted of two parts: First, the pretest was repeated, but with different variable values. Second, participants solved a programming task, which was to compute the average of all numbers between 1 and 100 that were multiples of 5 and print the result. The solutions were categorized into correct (2 points), conceptually mostly correct (1 point), and incorrect (0 points).

\paragraph{Interviews}
Within two weeks of the study, we conducted retrospective voluntary interviews with 4 participants from the course. Of these, 3 belonged to the \emph{Brocanto} group and 1 to the Git group. With the interviews, we gain detailed insights into possible applied strategies of the students in learning the artificial language as well as programming by having them reflect on the experiment, the Git introduction, and the course assignments. The interviews lasted about 30~minutes.
The participants received a small expense allowance.
The two course instructors each interviewed one participant together in a semi-structured interview. In general, participants were encouraged to reflect freely on topics. Open questions about possible problems in programming, repeating the tasks in the experiment and Git, and whether the treatment was helpful or hindering for learning programming provided the frames for reflection. The interview was digitally recorded and the recording was transcribed and anonymized in a further step for data protection and analysis reasons. 


\subsection{Participants}
To investigate our research question, we need participants interested in programming, but with no to little prior experience with programming. 

Thus, we were looking for students enrolled in their first semester at the Computer Science department from the first author. Via information events, flyers, and the Web site of the computer science department, we offered a voluntary pre-course for first-year students, in which basics of programming are taught. The experiments were part of this programming course. The researchers conducting the programming course were not involved in the courses that students completed in first semester. Despite extensive advertising of the course, we received fewer registrations than planned. Additionally, we excluded students who indicated at least medium programming experience. Since the students had not yet started their studies at that time, we were not able to contact all first semester students. We also could not reach any further participants after the course had started. With the interviews conducted after the course, we gathered more qualitative information to counter the small sample size.

In the end, 20 students participated in the study. We divided them into two balanced groups according to their prior programming experience and age, to ensure that both groups have comparable prerequisites for learning programming (cf.~Table~\ref{tab_group_division}). 
11 participants (8 men; 3 women, mean age 22 years) were in the \emph{Brocanto} group, and 9 participants (8 men, 1 woman, mean age 21 years) were in the control group. According to the pretest, both groups are comparable regarding their programming competency (\emph{Brocanto} group: 4.56 $\pm$ 3.05; Git group: 5.0 $\pm$ 3.08). In the end, however, 5 participants from the control group and 8 from the \emph{Brocanto} group took part in the posttest. Of these, 3 students of the control group completed the \emph{Algorithms and Programming} course, and 2 students of the treatment group.

\begin{table}
\caption{Group division according to pretest}
\centering
\begin{tabular}{lll|lll}
\toprule
Group     & Age & Gender & Group & Age & Gender  \\
\midrule
Treatment & 19  & m              & Control & 22 & m  \\
Treatment & 25  & m              & Control & 18 & m  \\
Treatment & 28  & m              & Control & 18 & m  \\
Treatment & 18  & m              & Control & 18 & m  \\
Treatment & 32  & w              & Control & 24 & m  \\
Treatment & 21  & m              & Control & 29 & m  \\
Treatment & 18  & w              & Control & 18 & m  \\
Treatment & 18  & m              & Control & 18 & m  \\
Treatment & 19  & m              & Control & 20 & w  \\
Treatment & 19  & m              \\
Treatment & 22  & w              \\
\bottomrule
\end{tabular}
\label{tab_group_division}
\end{table}

\subsection{Procedure}

We show an overview of the study in Figure~\ref{fig:Programming Course}. All 20 participants came to the same room, were greeted, and the details of the course were explained, after which the pretest and questionnaire were administered. On the next day, the \emph{Brocanto} group and Git group met in different rooms and learned \emph{Brocanto} or Git. The remaining days covered the introduction to programming, and the last day concluded with the posttest, after which the participants left. We conducted the interviews in the subsequent weeks. Six months after the experiment, we obtained the points that students achieved in their mandatory \emph{Algorithms and Programming} course.

\subsection{Deviations}

One of the participants mistakenly joined the \emph{Brocanto} group on the treatment day. That is why the number of participants is not perfectly balanced to 10 students per group; nevertheless, the groups have comparable programming competency according to the pretest. Furthermore, several participants dropped out because the course was too easy for them. This considerably reduces the power of our experiment.

\section{Results}

\subsection{Learning Brocanto}
\begin{figure}
    \centering
    \includegraphics[width = 0.5 \textwidth]{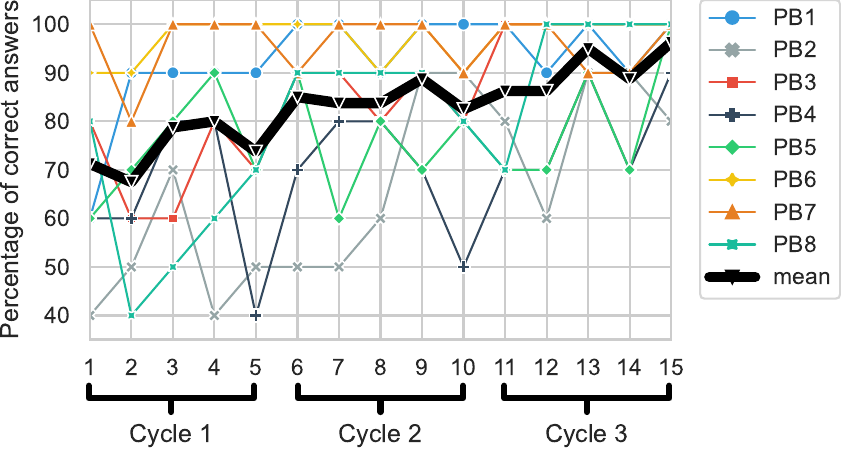}
    \caption{Correctness of the \emph{Brocanto} group in the test blocks. Note that the y-axis starts at 35.}
    \label{brocanto_lineplot}
\end{figure}

First, we evaluate whether participants learned \emph{Brocanto}. In Figure~\ref{brocanto_lineplot}, we show the performance of the \emph{Brocanto} group in the test blocks, in which they decided whether a sentence was correct or not. Correctness generally increases, so we can assume that the proficiency in \emph{Brocanto} has increased. Although not perfectly correct, these numbers are in line with typical studies, so we can conclude that students successfully learned \emph{Brocanto}~\cite{opitz2011timing, opitz2004brain}.

\subsection{Programming Competency}

\begin{figure*}[htbp]
\includegraphics[width = \textwidth]{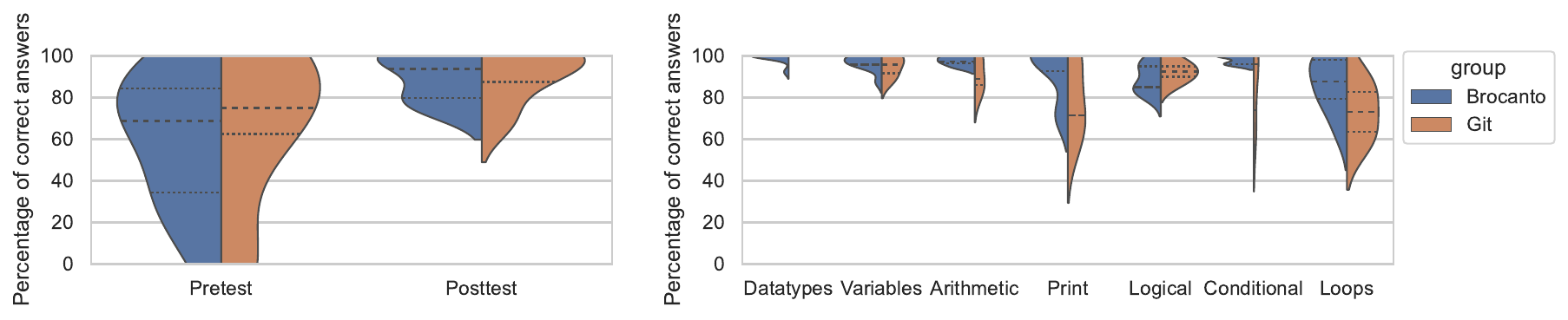}
\caption{Results of the pretest and posttest (left), as well as the intermediate assignments (right).}
\label{violinplot}
\end{figure*}

We evaluated the correctness of the programming tasks manually and generated descriptive and inferential statistics using the Python library Pandas. In Table~\ref{tab_mean_std} and Figure~\ref{violinplot}, we show an overview of the correctness for the pretest, intermediate assignments in the course, and the posttest.
The correctness is comparable across all tasks, but for completeness, we conducted a significance test. Since the assumptions for an ANOVA are not met (i.e., normal distribution), we use the Kruskal-Wallis test as non-parametric alternative. As expected with these comparable correctness scores and small sample size, none of the differences are significant.
However, it is interesting to note that the \emph{Brocanto} group performed better on the assignments of loops and conditionals. In both assignments, the \emph{Brocanto} group has an about 14\,\% higher correctness rate. This is an avenue for future work, either because learning these constructs is more closely linked to language learning strategies, or because these appear at the end of a course, hinting at a delayed effect of learning \emph{Brocanto}.
    
Thus, we can provide an answer to RQ\_2, that learning \emph{Brocanto} does not have an immediate positive effect on learning programming.
However, keeping in mind the small time frame of our observation, we might observe an effect at a later time. \\

For this reason, we took a look at the results of the post-programming tasks of their first-semester course. In the semester course \emph{Algorithms and Programming}, a total of 5 students, who also attended our course, continuously submitted at least 4 of the 7 programming tasks. The 3 participants from the control group scored 42 (PG2), 41 (PG4), and 11 (PG5) out of a total of 50 points, and the 2 participants from the \emph{Brocanto} group scored 28 (PB6) and 34 (PB8) points. To make a proper assumption on the long-term effect of the treatment, this data sample is too small. Nevertheless, the interview data provides us with some interesting insights, which we discuss in section \ref{Discussion} and which motivates us to pursue this line of research further.

Additionally, we can answer RQ\_1 positively, so we could feasibly integrate learning \emph{Brocanto} into our curriculum.
Our results indicate that considering Language Learning in a beginner programming course is justified. The observation that students were all able to learn \emph{Brocanto} suggests that there are "universal" learning strategies that students possess at least in the context of Language Learning. We hope that, in the future, recruiting more students for this or a similar course and including more long-term data will evaluate the (long-term) effects of beginning programming by learning an artificial language.

Next, we discuss our findings and insights from the interviews and derive conjectures for future research.

\section{Discussion}
\label{Discussion}
While our study did not show a significant improvement in programming learning, it has demonstrated that the experiment setup works well. 

The reason for the lacking significance could be the short time period between treatment and posttest. To observe the long-term development of the students, we obtained the scores from the programming tasks in the course \emph{Algorithms and Programming}, but due to a high drop out rate, we cannot draw sound conclusions from these data.

Against this background, we start the discussion by highlighting the feasibility of our approach and follow with learning strategies and possible consequences for learning programming. Further on, we explore whether transfer occurred. Finally, we discuss further insights that we gained from the study, that is, whether implicit or explicit teaching approaches affect learning strategies and the different levels of programming languages vs.\ natural languages.

\subsection{Does Learning \emph{Brocanto} Help in Learning Programming?}

In the interview, we asked participants how learning \emph{Brocanto} or Git affected them in learning programming. In general, participants considered learning \emph{Brocanto} neither helpful nor harmful: ``I don't know if it really helped or not. I'm not sure about that.''\footnote{All statements are translated to English.} 
(PB1), or ``So hindering in no way. To what extent it was helpful, I don't think I can judge that well, but it was definitely interesting.'' (PB5). For the git group we observe similar statements: ``But I don't think it helped in the course in particular, because we didn't do anything with Git and we only got a small introduction to Git.'' (PG5).

In general, participants agreed that learning an artificial language was a good idea to prepare them for learning programming. For example, PB3 explains that they would perform better in a repetition of this experiment at a later time, because ``then it would already be something familiar in a certain way, with this way of thinking, I say'' (PB3). When asked what is meant by ``this way of thinking'', PB3 specifies: ``so just this principle of looking at this [syntax] and learning it as fast as you can'' (PB3). This way of thinking is also mentioned by PB5, who describes a ``capacity for abstraction'' (PB5), where one ``first basically gets the idea of learning some grammar that doesn't directly serve to a language that you would converse with or something, but that is a bit more abstract and still understand the rules.'' (PB5). This statement suggests that participants apply abstraction, which is one (language) learning strategy~\cite{gick1983schema}.

Thus, there are hints of reactivation of learning strategies and that the participants assume, even 2 weeks after the implementation of the treatment, that they would perform better if they were to do it again. Thus, learning has taken place, and there is evidence that these structures could be used again.

\thesis{Learning \emph{Brocanto} activates (language) learning strategies, so it might help in learning programming.}

\subsection{Strategies in Learning Brocanto and Python}
\label{sec:Learning_Strategy}

There are indications of a systematic learning process of \emph{Brocanto}, such as the identification of structures and/or phrases of \emph{Brocanto} sentences. We observed four levels of abstraction: orientation, recognizing sentence structure, recognizing phrase structures, and recognizing words. We discuss these based on the interview data.

\subsubsection{Orientation}
PB1 generally points out that, when starting to learn the grammar, they have to ``orient themselves first'' (PB1), and PB3 mentioned that they ``[…] was looking at that and just really focused on looking at that and reading through it a couple of times, kind of lightly whispering to myself like that''. Thus, participants start with a general orientation phase, which is typical in learning languages to recognize structures~\cite{carrell1989metacognitive}.

\subsubsection{Recognizing Sentence Structure}
Not only the analysis of the syntax rules occurred, but also the division of sentences into ``beginning'', ``middle'', and the part ``following'' the middle, which is in line with research on artificial grammar learning~\cite{reber1967implicit, reber1989implicit}. PB5 explains: ``the [words] […] were always either at the beginning or after this part in the middle with such a word [consisting of] four letters […] \textsf{prez} I think and there was another that was relatively similar to that. And then there were two words that could then come directly after it and again two that then came partly as the third, whereby the one in the middle was sometimes also omitted.'' (PB5).

\subsubsection{Recognizing Phrase Structure}
PB3 describes an identification of phrase structures: ``you have in some places often such a three-word constructs or so, I don't know whether there was a rhythm, but there were just many sentences that then so always started with three words and then almost all sentences with these three words in the program or whatever. And those words that were in those groups, I kind of didn't memorize that well.'' (PB3). This is interesting linguistically, since in language learning, phrase-like sections of words, so-called chunks, can be used for orientation at the syntax level, assuming specifically three-element chunks ~\cite{franco2012chunking, meulemans1997associative, perruchet1990synthetic}. In a similar line, PB5 speaks of ``certain word combinations, [...] which then stood one after the other relatively frequently.'' (PB5). From this, patterns seem to have been derived: ``So that somehow gave the impression that they somehow belonged together or could somehow occur in a certain order.'' (PB5).

\subsubsection{Recognizing Words}
We observed a frequency effect that is also known from language learning: The frequency of occurrence (of a word or construct) increases the speed of its processing and recognition~\cite{ellis2002reflections, anderson2000adaptive}. Typically, words that appear more often are easier learned, as PB5 noted: ``[B]ecause you saw them the most. So they […] were even always there and so that was the most frequent repetition. The rest I think changed more than the ones at the beginning.'' (PB5). Interestingly, PB3 was better able to remember rare words: ``[I remember two words from artificial language,] Because they were rare, \textsf{füme}, I don't know how rare, but \textsf{plox} was rare. And the other words, the \textsf{gummen} (\textsf{gum}) I think there were more frequent.'' (PB3). In other words, the interruption of patterns by rarer words guided the structuring of syntax for extracting syntax rules.

Thus, participants apply typical language learning strategies to learn \emph{Brocanto}. Next, we discuss whether these strategies also come into play when learning Python.

\subsubsection{Language Learning Strategies When Learning Python}

In the interviews, we found several hints that these strategies actually play a role. First, there is a general orientation phase at the syntax level, because ``one has just recognized relatively quickly, with \emph{for} or with a \emph{for-loop}, what belongs in the range, and that it is just such a construct, which then probably also recognizes the IDE, so because of aha, so a \emph{for-loop} and so.'' (PB3). PB3 explains the orientation here to the span from the beginning of the \emph{for-loop} to the end. Again, we see an orientation to markers in the code. 

Similar to dividing sentences into ``beginning'', ``middle'', and ``following'' while learning \emph{Brocanto}, participants use Python markers as orientation to break down the parts into individual structures, as PB3 describes: 
``And also relatively logical (is) the indentation after \emph{if} and \emph{for} […]. I like that because first of all you see it directly and secondly, that's how I work too, when I make notes […] or something, then I just write how much of what belongs under what else, and then I move that in. That's logical and simple.'' (PB3) The statement ``see it directly'' refers to the graphical division of the individual parts of \emph{if-conditions} and \emph{for-loops} in Python. In \emph{Brocanto}, this part must be opened up through recognizing sentence structures (i.e., the second identified abstraction level). The indentation in Python makes this division visible, which facilitates learning; thus, a strategy does not have to applied explicitly, but is provided by the indentation. PB3 describes this: ``[Patterns are] the sequences, how you have to write it, for example, […] the loops. Especially that you have to indent and not use parentheses. And with print, for example, that always has to go in the parenthesis, about that order'' (PB1). Thus, indentation helps to identify sentence or phrase structures.

We also note that Python is seen here in comparison to other  programming languages that require parentheses instead of indentation. The parentheses of individual structures seem less helpful for orientation in the code and structuring of the individual programming constructs. Thus, in a programming language other than Python that would not follow this known pattern, structuring strategies might play a more important role. Thus, activating these strategies might be more important in programming languages without indentation.

We capture this observation in the following conjecture:

\thesis{Language learning strategies are used in learning programming.}
    
\subsection{Transfer}
Transfer describes that knowledge or skills from one domain are applied to a new domain. We found that transfer occurred from a programming language to Python and from a natural language to Python, but not from \emph{Brocanto} to Python.

\subsubsection{Programming Language $\rightarrow$ Python}

The interview data suggests that a transfer between programming languages takes place, providing further empirical support for recent studies by Shrestha~\cite{shrestha2020, shrestha2018s} and Tshukudu~\cite{tshukudu2020understanding}. A student with little previous contact to programming (in JavaScript and Java) explains that they had to pay attention to differences in syntax when learning to program in Python: ``And also before, I have always programmed with Java, I think. And there, the syntax is a bit different, especially with \emph{if-else}, there are no colons or there were some other things. In other programming languages, you have to put semicolons at the end, that's not the case with Python. Yes, so I had to be a bit careful that I don't forget or add things like that'' (PG5). On another point, the same participant states that differences between a learned programming language and a programming language to be learned can cause irritation in the learning process: ``Yes, that \emph{true} and \emph{false} are capitalized. That's not necessarily always like this [...] I find that irritating, because I would probably forget to write it in capital letters, because in other programming languages it doesn't even matter whether you write things in capital or small letters, and in Python it seems to matter. And I find that irritating '' (PG5). These statements show that, as in the transfer of natural languages, knowledge of the syntax is transferred to the new programming language. Differences between the source and target language involve the risk of transferring information incorrectly and thus making mistakes, but similarities can be transferred more quickly.

\thesis{Training of the differences and similarities between the source and target programming language facilitates transfer.}

\subsubsection{Natural Language $\rightarrow$ Python}

The structure of Python was generally rated as ``relatively logical'' (PG2) and not difficult by participants from the Git group as well as from the \emph{Brocanto} group. This impression was created among the participants, because they felt that Python was similar to natural languages, so ``very pleasant, because you had to pay attention to very few things. So, it was relative, I want to do that, I write it down and it fits like that.'' (PB3) The absent brackets also seemed to be beneficial: ``So, without, 'there must be five brackets here and there and so on.' […]'' (PB3), and ``as some of the syntax is not yet [understood], like these brackets that you have to think about […], you might understand it (Python) a little better at the beginning because it's closer to natural language […], or there aren't so many formal things'' (PG5).

A further benefit in Python was indentation, which provides structuring, similar to literal speech in natural-language texts, as PG2 describes: ``logical, […] that you just somehow indent a condition, you would perhaps also, I don't know, make quotation marks if you now quote someone or something'' (PG2). PG5 has a similar consideration: ``From a purely technical writing point of view, in Python you would somehow indent things and start new lines. If one compares that now with a written language, one would set perhaps a punctuation mark'' (PG5).  Thus, it might be easier for students to start off with variants of programming or artificial languages that are close to natural languages and step by step become more similar to a full programming language. \emph{Hedy} is one example of starting close to natural languages, and step by step becoming Python~\cite{Hermans20}. 

\thesis{When learning a first programming language, a closeness to natural language is beneficial.}

\thesis{The more programming languages a person is familiar with, the farther away they can be from natural language.} 

However, there is also interference from natural language to Python, especially where natural languages and Python differ. Especially features in syntax have to be applied several times until it is learned. PB3 calls for this ``that in the beginning I always forgot to add the colon to \emph{if} or \emph{for}'' (PB3). A participant from the control group can also generally imagine that, in connection between natural and artificial languages, and Python, ``[learning an artificial language] might be a bit confusing, because some of the rules or typical structures of the natural language are still in there. And if you then use the programming language, it can perhaps happen that you still use these things'' (PG5).
Thus, despite the logical structure of Python, which can be logically linked to structures of natural languages, ``this structure, these strands, which you have to follow, and at the beginning especially the syntax in general'' (PG2) are considered as ``very difficult'' (PG2) when learning programming, which is in line with research describing that actually learning syntax is a struggle for students~\cite{denny_understanding_2011, edwards_syntax_2020}. These statements suggest that the differences in the states of and to which knowledge is to be transferred can lead to problems. We state

\thesis{When learning a new programming language, providing explicit guidance on similarities and differences is beneficial.}

\thesis{Providing focused guidance on difficult to memorize constructs supports learning programming.}

\noindent In the interviews, students focused on learning the syntax. However, one participant made an interesting statement about the semantics level, which also hints at a specific difference between natural language and programming languages, but also at why training with an artificial language may be helpful. In this statement, PG2 concerns ``going through loops'' (PG2) and the fact that one does not see on the written level of the program ``how the [program] does [the loop] now, but you have only written there, what is to be done and does not see however, how that is done, except you let it give the output.'' (PG2) The participant describes the semantic level, which cannot be decoded for programming beginners just by learning the syntax. At this level of acquisition, the code must be output to see what it ``does''. With natural languages, semantic understanding is already present from the first language learned, because people are familiar with the semantic system in natural languages. However, in programming, a new semantic system has to be learned, because it is different from natural languages. In this case, the artificial language can act as intermediate carrier, because it has a natural language syntax, but has no semantics in the strict sense. Thus, we conjecture that:

\thesis{Training with artificial languages serves as an intermediate stage for learning semantics of programming languages.} 

\subsection{Further Insights}

\subsubsection{Implicit vs.\ Explicit Teaching}

Some participants highlighted in the interview that they intuitively learned \emph{Brocanto}, without being taught explicitly the syntactic rules. This implicit learning of artificial languages is the typical protocol in experiments with artificial languages~\cite{opitz2011timing, perruchet1990synthetic, meulemans1997associative, kinder2000learning, reber1967implicit, reber1989implicit}, and often likened with gut feeling, or ``intuitive'', or ``spontaneous'' learning, or ``learning without awareness'': ``I honestly didn't know, I think, I felt like I had a relatively large amount right. […] I didn't know why, though.'' (PB3) The same participant states, they ``actively, have not detected any system'' (PB3). Thus, the participant was aware that they had not actively recognized and used any system, but had nevertheless improved in the correctness of the test blocks (red in Fig.~\ref{brocanto_lineplot}. Strikingly, this participant gave the definition of implicit learning~\cite{frensch2003implicit}: ``I just looked at [the sentences] briefly and then I just pressed the first thing that came into my mind. […] I just did it really intuitively'' (PB3). This ``gut feeling'' or implicit learning can also be measured objectively (e.g.,~\cite{reber1967implicit, reber1989implicit}): When syntactic violations occur in acquired \emph{Brocanto} sentences, event-related potentials for syntax violations can be observed, which are different than for correct sentences. Thus, the brain gives an objective signal of intuitive learning.

One participant of the control group explained that they generally had ``somehow been able to acquire the structures [of a programming language]'' (PG5). The assumption that this happened implicitly is supported by a statement of the same person at a later point in the interview. In this they denied that they had explicitly worked out structures as a learning strategy before, or applied them: ``I'm no such person [like the ones that] even in [school subject] […] structure something. I have never done that'' (PG5). This statement refers to a general implicit structuring of the programming language during learning. Individual aspects are not mentioned. 
However, it might also be observable for programming languages, that syntactically incorrect statements elicit a different neuronal response than correct statements, which might be a further hint to the similarity between natural, artificial, and programming languages, making the intermediate step of learning an artificial language viable. 

\thesis{Syntactic violations in statements in a programming language elicit a specific neuronal response.}

Interestingly, despite describing implicit learning, the participant described at a later point concrete rules from \emph{Brocanto}, indicating that they identified explicit rules (e.g., ``three-word constructs […]many sentences that always started with three words and then almost all sentences [started] with these three words.'' (PB3); cf.\ Section~\ref{sec:Learning_Strategy}). In contrast to the implicit learning of the artificial language, participants perceived the learning of Python as explicit: ``Because the [Python] rules were given and not just examples from which you could derive the rules, I think it was even easier to understand than with this artificial language.'' (PG2). Nevertheless, there are also hints that knowledge about a programming language can also be implicit~\cite{Mancy07}.
 
For each new programming construct, participants received templates for the respective construct, such that they can use them with the variables of their program and combine constructs in more complex tasks. Thus, the participants were explicitly instructed that these templates will be relevant in the further process of programming. When asked about the use of the template, PB3 states: ``I thought it was good in any case, […] that you always got the basics first, because of, this is how it looks, this is how it is built up from the […] (syntactic) components, is this structure put together, these are the rules, how do you write this and so on […]. For me, it definitely helped to [learn] the structure.'' Here, it becomes clear that the participant does not have to derive the rules themselves, but understands the given rules for syntactic structures in the programming language and then applies them. For the intermediate programming assignments and posttest, the participants then ``puts together everything […] that I learned'' (PT3). 

To summarize, \emph{Brocanto} was learned implicitly, Python explicitly. Explicit learning seems to make it easier for students to understand the rules, but implicit learning of \emph{Brocanto} is closer to natural language learning. Thus, to activate strategies from natural language learning and apply an explicit learning approach, a \emph{two-step} approach that starts with teaching an artificial language implicitly (to tap into the language system), and then teaches an(other) artificial language explicitly (to support recognition of rules) might prove effective.

\thesis{Combining implicit and explicit teaching approaches of an artificial language positively influences initial programming language learning.} 

\subsubsection{Meta-Level Semantics}

In the interview, participants reflected on problems that they had in learning programming. It describes the semantic level of code and of natural languages. It became clear that code has more levels than text. The principle behind code is described as ``very abstract'' (PG5). The intent of the program, that is, what it ``does'' (PG5), is not directly clear until the program runs and produces an output. PG5 gives an example: ``You have just letters or something, […] or variables that you remember as something or that you define and that you then have all the time and that just count like that.'' (PG5). As another example, even though a loop is written once, it is not necessarily executed once, but likely several times, depending on the concrete input. And with different input, the same loop can behave differently. Thus, what is written in code first needs to be re-written internally to truly reflect what it is doing.
This is different in natural languages, in which the intent of a sentence is already the sentence itself. The message to be conveyed is concretely conveyed by the sentence itself, so there is a closer connection between text and semantics. Thus, it may be helpful for beginner programmers to translate a piece of source code into its actual behavior and write it down, for example, all loop iterations (given reasonable input) or replace the occurrences of all variables with actual values. This might help to understand how the different level of semantic in code (i.e., its behavior) translates to the actual, written source code.

\thesis{Explicitly writing down the specific behavior of code helps to bridge the different levels of code.}
 
\subsection{Summary of Discussion}

The interviews provided valuable insights, including the activation of language learning strategies by training an artificial language, as well as new approaches to transfer, that is, transferring known patterns to a similar, yet unknown subject, from natural language to programming language or between programming languages.
In addition, we made observations regarding implicit and explicit learning, which concern new approaches of teaching programming languages as well as artificial languages. Furthermore, we highlighted the difficulty of beginners to decode the semantics behind the programming language. This differs at the comprehension level compared to the sentence-level semantics in natural languages. This difference seems to make it difficult to comprehend programming language constructs.\\

Even though these conjectures were derived from a small sample size and may depend on factors, such as personal preferences and existing programming and language learning experience, they give us interesting insights and inspiration that are worth wile to follow up on. With this new angle on programming learning, we hope that in the future, programming learning will become easier and more students are motivated to pursue it.

\section{Threats to Validity}

\subsection{Construct Validity}

To mitigate potential threats caused by inaccurate measurement of the latent constructs, we used an evaluated programming test, and also worked with a well-established questionnaire to measure programming experience. In the same line, we followed a standard protocol to teach \emph{Brocanto}, and followed a curriculum of teaching programming. Thus, threats to construct validity are minimized.

\subsection{Internal Validity}

The programming course took place within a week and thus in a short time frame to sufficiently observe the influence of the treatment on programming learning. To get an impression of longer-term effects, we looked at the results of the programming tasks from the first-semester course \emph{Algorithms and Programming} six months after the course. Due to the high drop-out rate in the course, we can only compare a few results.

The two very similar groups differed only in the treatment, so we can take a good look at the effect of the treatment. In our study, we included participants with no programming experience as well as participants with little programming experience. We could not assess the extent to which prior experience might influence our research. However, data from the first-semester course \emph{Algorithms and Programming} show that participants from our course who reported having some programming experience did not necessarily perform better in the programming tasks than participants without programming experience. Thus, we count them as beginners as the participants who had no programming experience.
It is quite possible that at the measuring points, despite the request to work on the tests alone, the tests were worked on with the person sitting next to them or were larded. Since, in addition to the treatment, all participants took part in the programming course and all significantly improved their programming ability, it is possible to conclude that the course had an effect on performance. The effect of the treatment can be separated, but not observed in the results.
The control treatment of the Git group did not include any programming aspects. Nevertheless, we cannot fully rule out the possibility that further engagement with Git taught strategies that may have contributed to learning programming.

Furthermore, some students dropped out, because it was too easy for them. This can also be seen in the 100\,\% correctness in their pretest scores. While this reduces statistical power, it does not threaten internal validity, as the too high programming competency might have confounded the results more in favor for the control group.

\subsection{External Validity}

In the course, we worked only with Python and \emph{Brocanto}, and covered only one week. With another programming and/or artificial language, or courses spanning longer time spans, results may be different.
In addition, all of the students were part of the computer science faculty, so a general interest in computer science already existed. We cannot say with certainty whether students from other departments would learn the artificial language or programming language with different strategies. Nevertheless, our results are applicable to this important population of students who learn Python or similar programming languages.

\section{Conclusion}

Teaching the first programming language is accompanied by many difficulties. Research has shown that pairing learning programming with foreign language acquisition can be beneficial. In conjunction, activating language learning strategies can facilitate the learning of the first programming language through transfer. In our study as part of a programming course, we investigated whether students who learn an artificial language before learning programming better acquire programming competency. 
Within the time frame of our study, we did not find a general effect on programming learning when an artificial language was previously learned. However, we observed that the participants learning the artificial language performed better on the intermediate assignments on conditionals and loops, which may indicate that strategies involved in the structuring that were used in learning the artificial language. However, we cannot derive a clear result yet. We also took a long-term look at programming outcomes from the students' first semester; but due to the high drop-out rates, could not make any estimates of the long-term effects of the treatment. The conducted study will serve as a basis for the new programming course in the next winter semester. \\

On a methodological level, we learned that experiment designs from the field of foreign language acquisition research can be used for fathoming computer science problems. By doing so, we represent a step on the road map that can be explored for computer science research. We highlighted relevant points on the roadmap by providing \numthesis conjectures that are good starting points to explore this experiment landscape. 

\section*{Acknowledgment}

We thank the students for participating in the study. Apel's work is supported by ERC Advanced Grant 101052182.

\bibliographystyle{ACM-Reference-Format}
\bibliography{bibliography}


\begin{thebibliography}{51}


\ifx \showCODEN    \undefined \def \showCODEN     #1{\unskip}     \fi
\ifx \showDOI      \undefined \def \showDOI       #1{#1}\fi
\ifx \showISBNx    \undefined \def \showISBNx     #1{\unskip}     \fi
\ifx \showISBNxiii \undefined \def \showISBNxiii  #1{\unskip}     \fi
\ifx \showISSN     \undefined \def \showISSN      #1{\unskip}     \fi
\ifx \showLCCN     \undefined \def \showLCCN      #1{\unskip}     \fi
\ifx \shownote     \undefined \def \shownote      #1{#1}          \fi
\ifx \showarticletitle \undefined \def \showarticletitle #1{#1}   \fi
\ifx \showURL      \undefined \def \showURL       {\relax}        \fi
\providecommand\bibfield[2]{#2}
\providecommand\bibinfo[2]{#2}
\providecommand\natexlab[1]{#1}
\providecommand\showeprint[2][]{arXiv:#2}

\bibitem[Ahadi et~al\mbox{.}(2014)]%
        {ahadi2014falling}
\bibfield{author}{\bibinfo{person}{Alireza Ahadi}, \bibinfo{person}{Raymond
  Lister}, {and} \bibinfo{person}{Donna Teague}.}
  \bibinfo{year}{2014}\natexlab{}.
\newblock \showarticletitle{{Falling Behind Early and Staying Behind When
  Learning to Program}}. In \bibinfo{booktitle}{\emph{PPIG}},
  Vol.~\bibinfo{volume}{14}. \bibinfo{numpages}{12}~pages.
\newblock


\bibitem[Anderson and Schooler(2000)]%
        {anderson2000adaptive}
\bibfield{author}{\bibinfo{person}{John Anderson} {and} \bibinfo{person}{Lael
  Schooler}.} \bibinfo{year}{2000}\natexlab{}.
\newblock \showarticletitle{{The Adaptive Nature of Memory}}.
\newblock  (\bibinfo{year}{2000}).
\newblock


\bibitem[Batterink and Neville(2013)]%
        {batterink2013implicit}
\bibfield{author}{\bibinfo{person}{Laura Batterink} {and}
  \bibinfo{person}{Helen Neville}.} \bibinfo{year}{2013}\natexlab{}.
\newblock \showarticletitle{{Implicit and Explicit Second Language Training
  recruit Common Neural Mechanisms for Syntactic Processing}}.
\newblock \bibinfo{journal}{\emph{Journal of Cognitive Neuroscience}}
  \bibinfo{volume}{25}, \bibinfo{number}{6} (\bibinfo{year}{2013}),
  \bibinfo{pages}{936--951}.
\newblock


\bibitem[Bennedsen and Caspersen(2007)]%
        {bennedsen_failure_2007}
\bibfield{author}{\bibinfo{person}{Jens Bennedsen} {and}
  \bibinfo{person}{Michael Caspersen}.} \bibinfo{year}{2007}\natexlab{}.
\newblock \showarticletitle{{Failure Rates in Introductory Programming}}.
\newblock \bibinfo{journal}{\emph{ACM SIGCSE Bulletin}} \bibinfo{volume}{39},
  \bibinfo{number}{2} (\bibinfo{year}{2007}), \bibinfo{pages}{32--36}.
\newblock


\bibitem[Bennedsen and Caspersen(2019)]%
        {bennedsen_failure_2019}
\bibfield{author}{\bibinfo{person}{Jens Bennedsen} {and}
  \bibinfo{person}{Michael Caspersen}.} \bibinfo{year}{2019}\natexlab{}.
\newblock \showarticletitle{{Failure Rates in Introductory Programming: 12
  Years later}}.
\newblock \bibinfo{journal}{\emph{ACM Inroads}} \bibinfo{volume}{10},
  \bibinfo{number}{2} (\bibinfo{year}{2019}), \bibinfo{pages}{30--36}.
\newblock


\bibitem[Brooks and Vokey(1991)]%
        {brooks1991abstract}
\bibfield{author}{\bibinfo{person}{Lee Brooks} {and} \bibinfo{person}{John
  Vokey}.} \bibinfo{year}{1991}\natexlab{}.
\newblock \showarticletitle{{Abstract Analogies and Abstracted Grammars:
  Comments on Reber (1989) and Mathews et al. (1989).}}
\newblock  (\bibinfo{year}{1991}).
\newblock


\bibitem[Carrell(1989)]%
        {carrell1989metacognitive}
\bibfield{author}{\bibinfo{person}{Patricia Carrell}.}
  \bibinfo{year}{1989}\natexlab{}.
\newblock \showarticletitle{{Metacognitive Awareness and Second Language
  Reading}}.
\newblock \bibinfo{journal}{\emph{The modern language journal}}
  \bibinfo{volume}{73}, \bibinfo{number}{2} (\bibinfo{year}{1989}),
  \bibinfo{pages}{121--134}.
\newblock


\bibitem[Cooper(2019)]%
        {cooper2019influence}
\bibfield{author}{\bibinfo{person}{Alison Cooper}.}
  \bibinfo{year}{2019}\natexlab{}.
\newblock \bibinfo{booktitle}{\emph{{The Influence of Position in the Sleep
  Wake Cycle on the Functional Brain Processes involved in Memory Consolidation
  during Acquisition of an Artificial Language (BROCANTO)}}}.
\newblock \bibinfo{publisher}{University of Surrey (United Kingdom)}.
\newblock


\bibitem[Denny et~al\mbox{.}(2011)]%
        {denny_understanding_2011}
\bibfield{author}{\bibinfo{person}{Paul Denny}, \bibinfo{person}{Andrew
  Luxton-Reilly}, \bibinfo{person}{Ewan Tempero}, {and} \bibinfo{person}{Jacob
  Hendrickx}.} \bibinfo{year}{2011}\natexlab{}.
\newblock \showarticletitle{Understanding the Syntax Barrier for Novices}. In
  \bibinfo{booktitle}{\emph{Proc. Conf. {Innovation} and Technology in Computer
  Science Education}} \emph{(\bibinfo{series}{{ITiCSE} '11})}.
  \bibinfo{publisher}{Association for Computing Machinery},
  \bibinfo{address}{New York, NY, USA}, \bibinfo{pages}{208--212}.
\newblock


\bibitem[Edwards et~al\mbox{.}(2020)]%
        {edwards_syntax_2020}
\bibfield{author}{\bibinfo{person}{John Edwards}, \bibinfo{person}{Joseph
  Ditton}, \bibinfo{person}{Dragan Trninic}, \bibinfo{person}{Hillary Swanson},
  \bibinfo{person}{Shelsey Sullivan}, {and} \bibinfo{person}{Chad Mano}.}
  \bibinfo{year}{2020}\natexlab{}.
\newblock \showarticletitle{Syntax {Exercises} in {CS1}}. In
  \bibinfo{booktitle}{\emph{Proc. {Conf.} on {International} {Computing}
  {Education} {Research}}} \emph{(\bibinfo{series}{{ICER} '20})}.
  \bibinfo{publisher}{ACM}, \bibinfo{address}{New York, NY, USA},
  \bibinfo{pages}{216--226}.
\newblock


\bibitem[Ellis(2002)]%
        {ellis2002reflections}
\bibfield{author}{\bibinfo{person}{Nick Ellis}.}
  \bibinfo{year}{2002}\natexlab{}.
\newblock \showarticletitle{{Reflections on Frequency Effects in Language
  Processing}}.
\newblock \bibinfo{journal}{\emph{Studies in second language acquisition}}
  \bibinfo{volume}{24}, \bibinfo{number}{2} (\bibinfo{year}{2002}),
  \bibinfo{pages}{297--339}.
\newblock


\bibitem[Ettlinger et~al\mbox{.}(2016)]%
        {ettlinger2016relationship}
\bibfield{author}{\bibinfo{person}{Marc Ettlinger}, \bibinfo{person}{Kara
  Morgan-Short}, \bibinfo{person}{Mandy Faretta-Stutenberg}, {and}
  \bibinfo{person}{Patrick Wong}.} \bibinfo{year}{2016}\natexlab{}.
\newblock \showarticletitle{{The Relationship between Artificial and Second
  Language Learning}}.
\newblock \bibinfo{journal}{\emph{Cognitive science}} \bibinfo{volume}{40},
  \bibinfo{number}{4} (\bibinfo{year}{2016}), \bibinfo{pages}{822--847}.
\newblock


\bibitem[Franco and Destrebecqz(2012)]%
        {franco2012chunking}
\bibfield{author}{\bibinfo{person}{Ana Franco} {and} \bibinfo{person}{Arnaud
  Destrebecqz}.} \bibinfo{year}{2012}\natexlab{}.
\newblock \showarticletitle{{Chunking or Not Chunking? How do we find Words in
  Artificial Language Learning?}}
\newblock \bibinfo{journal}{\emph{Advances in Cognitive Psychology}}
  \bibinfo{volume}{8}, \bibinfo{number}{2} (\bibinfo{year}{2012}),
  \bibinfo{pages}{144}.
\newblock


\bibitem[Frensch and R{\"u}nger(2003)]%
        {frensch2003implicit}
\bibfield{author}{\bibinfo{person}{Peter Frensch} {and} \bibinfo{person}{Dennis
  R{\"u}nger}.} \bibinfo{year}{2003}\natexlab{}.
\newblock \showarticletitle{{Implicit Learning}}.
\newblock \bibinfo{journal}{\emph{Current Directions in Psychological Science}}
  \bibinfo{volume}{12}, \bibinfo{number}{1} (\bibinfo{year}{2003}),
  \bibinfo{pages}{13--18}.
\newblock


\bibitem[Friederici et~al\mbox{.}(2002)]%
        {friederici2002brain}
\bibfield{author}{\bibinfo{person}{Angela Friederici}, \bibinfo{person}{Karsten
  Steinhauer}, {and} \bibinfo{person}{Erdmut Pfeifer}.}
  \bibinfo{year}{2002}\natexlab{}.
\newblock \showarticletitle{{Brain Signatures of Artificial Language
  Processing: Evidence Challenging the Critical Period Hypothesis}}.
\newblock \bibinfo{journal}{\emph{Proc. of the National Academy of Sciences}}
  \bibinfo{volume}{99}, \bibinfo{number}{1} (\bibinfo{year}{2002}),
  \bibinfo{pages}{529--534}.
\newblock


\bibitem[Gauthier et~al\mbox{.}(2000)]%
        {Gauthier00}
\bibfield{author}{\bibinfo{person}{Isabel Gauthier}, \bibinfo{person}{Pawel
  Skudlarski}, \bibinfo{person}{John Gore}, {and} \bibinfo{person}{Adam
  Anderson}.} \bibinfo{year}{2000}\natexlab{}.
\newblock \showarticletitle{{Expertise for Cars and Birds Recruits Brain Areas
  Involved in Face Recognition}}.
\newblock \bibinfo{journal}{\emph{Nature Neuroscience}} \bibinfo{volume}{3},
  \bibinfo{number}{2} (\bibinfo{date}{Feb} \bibinfo{year}{2000}),
  \bibinfo{pages}{191--197}.
\newblock


\bibitem[Gick and Holyoak(1983)]%
        {gick1983schema}
\bibfield{author}{\bibinfo{person}{Mary Gick} {and} \bibinfo{person}{Keith
  Holyoak}.} \bibinfo{year}{1983}\natexlab{}.
\newblock \showarticletitle{{Schema Induction and Analogical Transfer}}.
\newblock \bibinfo{journal}{\emph{Cognitive psychology}} \bibinfo{volume}{15},
  \bibinfo{number}{1} (\bibinfo{year}{1983}), \bibinfo{pages}{1--38}.
\newblock


\bibitem[Hermans(2020)]%
        {Hermans20}
\bibfield{author}{\bibinfo{person}{Felienne Hermans}.}
  \bibinfo{year}{2020}\natexlab{}.
\newblock \showarticletitle{{Hedy: A Gradual Language for Programming
  Education}}. In \bibinfo{booktitle}{\emph{Int.\ Conf.\ Educational Research
  (ICER)}}. \bibinfo{publisher}{ACM}, \bibinfo{pages}{259--270}.
\newblock


\bibitem[Hongo et~al\mbox{.}(2022)]%
        {Hongo22}
\bibfield{author}{\bibinfo{person}{Takeshi Hongo}, \bibinfo{person}{Takao
  Yakou}, \bibinfo{person}{Kenji Yoshinaga}, \bibinfo{person}{Toshiharu Kano},
  \bibinfo{person}{Michiko Miyazaki}, {and} \bibinfo{person}{Takashi
  Hanakawa}.} \bibinfo{year}{2022}\natexlab{}.
\newblock \showarticletitle{{{Structural Neuroplasticity in Computer
  Programming Beginners}}}.
\newblock \bibinfo{journal}{\emph{Cerebral Cortex}} (\bibinfo{date}{10}
  \bibinfo{year}{2022}).
\newblock
\newblock
\shownote{Online first}.


\bibitem[Kinder and Assmann(2000)]%
        {kinder2000learning}
\bibfield{author}{\bibinfo{person}{Annette Kinder} {and} \bibinfo{person}{Anja
  Assmann}.} \bibinfo{year}{2000}\natexlab{}.
\newblock \showarticletitle{{Learning Artificial Grammars: No Evidence for the
  Acquisition of Rules}}.
\newblock \bibinfo{journal}{\emph{Memory \& Cognition}} \bibinfo{volume}{28},
  \bibinfo{number}{8} (\bibinfo{year}{2000}), \bibinfo{pages}{1321--1332}.
\newblock


\bibitem[Knowlton and Squire(1996)]%
        {knowlton1996artificial}
\bibfield{author}{\bibinfo{person}{Barbara Knowlton} {and}
  \bibinfo{person}{Larry Squire}.} \bibinfo{year}{1996}\natexlab{}.
\newblock \showarticletitle{{Artificial Grammar Learning Depends on Implicit
  Acquisition of both Abstract and Exemplar-Specific Information}}.
\newblock \bibinfo{journal}{\emph{Journal of Experimental Psychology: Learning,
  Memory, and Cognition}} \bibinfo{volume}{22}, \bibinfo{number}{1}
  (\bibinfo{year}{1996}), \bibinfo{pages}{169}.
\newblock


\bibitem[Krueger et~al\mbox{.}(2020)]%
        {krueger2020neurological}
\bibfield{author}{\bibinfo{person}{Ryan Krueger}, \bibinfo{person}{Yu Huang},
  \bibinfo{person}{Xinyu Liu}, \bibinfo{person}{Tyler Santander},
  \bibinfo{person}{Westley Weimer}, {and} \bibinfo{person}{Kevin Leach}.}
  \bibinfo{year}{2020}\natexlab{}.
\newblock \showarticletitle{{Neurological Divide: an fMRI Study of Prose and
  Code Writing}}. In \bibinfo{booktitle}{\emph{Proc.\ Int'l Conf.\ Software
  Engineering (ICSE)}}. \bibinfo{publisher}{IEEE}, \bibinfo{pages}{678--690}.
\newblock


\bibitem[Lister et~al\mbox{.}(2004)]%
        {lister_multi-national_2004}
\bibfield{author}{\bibinfo{person}{Raymond Lister},
  \bibinfo{person}{Elizabeth~S. Adams}, \bibinfo{person}{Sue Fitzgerald},
  \bibinfo{person}{William Fone}, \bibinfo{person}{John Hamer},
  \bibinfo{person}{Morten Lindholm}, \bibinfo{person}{Robert McCartney},
  \bibinfo{person}{Jan~Erik Moström}, \bibinfo{person}{Kate Sanders},
  \bibinfo{person}{Otto Seppälä}, \bibinfo{person}{Beth Simon}, {and}
  \bibinfo{person}{Lynda Thomas}.} \bibinfo{year}{2004}\natexlab{}.
\newblock \showarticletitle{{A Multi-National Study of Reading and Tracing
  Skills in Novice Programmers}}.
\newblock \bibinfo{journal}{\emph{ACM SIGCSE Bulletin}} \bibinfo{volume}{36},
  \bibinfo{number}{4} (\bibinfo{year}{2004}), \bibinfo{pages}{119--150}.
\newblock
\showISSN{0097-8418}


\bibitem[Lister et~al\mbox{.}(2006)]%
        {lister_asssesment_2006}
\bibfield{author}{\bibinfo{person}{Raymond Lister}, \bibinfo{person}{Beth
  Simon}, \bibinfo{person}{Errol Thompson}, \bibinfo{person}{Jacqueline
  Whalley}, {and} \bibinfo{person}{Christine Prasad}.}
  \bibinfo{year}{2006}\natexlab{}.
\newblock \showarticletitle{{Not Seeing the Forest for the Trees: Novice
  Programmers and the SOLO Taxonomy}}.
\newblock \bibinfo{journal}{\emph{SIGCSE Bull.}} \bibinfo{volume}{38},
  \bibinfo{number}{3} (\bibinfo{year}{2006}), \bibinfo{pages}{118–122}.
\newblock


\bibitem[Luxton-Reilly(2016)]%
        {luxton-reilly_learning_2016}
\bibfield{author}{\bibinfo{person}{Andrew Luxton-Reilly}.}
  \bibinfo{year}{2016}\natexlab{}.
\newblock \showarticletitle{Learning to {Program} is {Easy}}. In
  \bibinfo{booktitle}{\emph{Proc. {Conf.} {Innovation} and {Technology} in
  {Computer} {Science} {Education}}}. \bibinfo{publisher}{ACM},
  \bibinfo{address}{Arequipa Peru}, \bibinfo{pages}{284--289}.
\newblock


\bibitem[Mancy(2007)]%
        {Mancy07}
\bibfield{author}{\bibinfo{person}{Rebecca Mancy}.}
  \bibinfo{year}{2007}\natexlab{}.
\newblock \emph{\bibinfo{title}{{Explicit and Implicit Learning: The Case of
  Computer Programming}}}.
\newblock \bibinfo{thesistype}{Ph.\,D. Dissertation}.
  \bibinfo{school}{University of Glasgow}.
\newblock


\bibitem[McCracken et~al\mbox{.}(2001)]%
        {mccracken_multi-national_2001}
\bibfield{author}{\bibinfo{person}{Michael McCracken}, \bibinfo{person}{Vicki
  Almstrum}, \bibinfo{person}{Danny Diaz}, \bibinfo{person}{Mark Guzdial},
  \bibinfo{person}{Dianne Hagan}, \bibinfo{person}{Yifat Ben-David Kolikant},
  \bibinfo{person}{Cary Laxer}, \bibinfo{person}{Lynda Thomas},
  \bibinfo{person}{Ian Utting}, {and} \bibinfo{person}{Tadeusz Wilusz}.}
  \bibinfo{year}{2001}\natexlab{}.
\newblock \showarticletitle{{A Multi-National, Multi-Institutional Study of
  Assessment of Programming Skills of First-Year {CS} Students}}. In
  \bibinfo{booktitle}{\emph{Working group reports from {ITiCSE} on {Innovation}
  and technology in computer science education}}
  \emph{(\bibinfo{series}{{ITiCSE}-{WGR} '01})}.
  \bibinfo{publisher}{Association for Computing Machinery},
  \bibinfo{address}{New York, NY, USA}, \bibinfo{pages}{125--180}.
\newblock


\bibitem[Meulemans and Van~der Linden(1997)]%
        {meulemans1997associative}
\bibfield{author}{\bibinfo{person}{Thierry Meulemans} {and}
  \bibinfo{person}{Martial Van~der Linden}.} \bibinfo{year}{1997}\natexlab{}.
\newblock \showarticletitle{{Associative Chunk Strength in Artificial Grammar
  Learning.}}
\newblock \bibinfo{journal}{\emph{Journal of Experimental Psychology: Learning,
  Memory, and Cognition}} \bibinfo{volume}{23}, \bibinfo{number}{4}
  (\bibinfo{year}{1997}), \bibinfo{pages}{1007}.
\newblock


\bibitem[Opitz et~al\mbox{.}(2011)]%
        {opitz2011timing}
\bibfield{author}{\bibinfo{person}{Bertram Opitz}, \bibinfo{person}{Nicola
  Ferdinand}, {and} \bibinfo{person}{Axel Mecklinger}.}
  \bibinfo{year}{2011}\natexlab{}.
\newblock \showarticletitle{{Timing Matters: the Impact of Immediate and
  Delayed Feedback on Artificial Language Learning}}.
\newblock \bibinfo{journal}{\emph{Frontiers in human neuroscience}}
  \bibinfo{volume}{5} (\bibinfo{year}{2011}), \bibinfo{pages}{8}.
\newblock


\bibitem[Opitz and Friederici(2003)]%
        {opitz2003interactions}
\bibfield{author}{\bibinfo{person}{Bertram Opitz} {and} \bibinfo{person}{Angela
  Friederici}.} \bibinfo{year}{2003}\natexlab{}.
\newblock \showarticletitle{{Interactions of the Hippocampal System and the
  Prefrontal Cortex in Learning Language-Like Rules}}.
\newblock \bibinfo{journal}{\emph{NeuroImage}} \bibinfo{volume}{19},
  \bibinfo{number}{4} (\bibinfo{year}{2003}), \bibinfo{pages}{1730--1737}.
\newblock


\bibitem[Opitz and Friederici(2004)]%
        {opitz2004brain}
\bibfield{author}{\bibinfo{person}{Bertram Opitz} {and} \bibinfo{person}{Angela
  Friederici}.} \bibinfo{year}{2004}\natexlab{}.
\newblock \showarticletitle{{Brain Correlates of Language Learning: the
  Neuronal Dissociation of Rule-Based versus Similarity-Based Learning}}.
\newblock \bibinfo{journal}{\emph{Journal of Neuroscience}}
  \bibinfo{volume}{24}, \bibinfo{number}{39} (\bibinfo{year}{2004}),
  \bibinfo{pages}{8436--8440}.
\newblock


\bibitem[Peitek et~al\mbox{.}(2021)]%
        {ICSE21}
\bibfield{author}{\bibinfo{person}{Norman Peitek}, \bibinfo{person}{Sven Apel},
  \bibinfo{person}{Chris Parnin}, \bibinfo{person}{Andr\'e Brechmann}, {and}
  \bibinfo{person}{Janet Siegmund}.} \bibinfo{year}{2021}\natexlab{}.
\newblock \showarticletitle{{Program Comprehension and Code Complexity Metrics:
  An fMRI Study}}. In \bibinfo{booktitle}{\emph{ICSE}}.
  \bibinfo{publisher}{IEEE}.
\newblock


\bibitem[Perruchet and Pacteau(1990)]%
        {perruchet1990synthetic}
\bibfield{author}{\bibinfo{person}{Pierre Perruchet} {and}
  \bibinfo{person}{Chantal Pacteau}.} \bibinfo{year}{1990}\natexlab{}.
\newblock \showarticletitle{Synthetic Grammar Learning: Implicit Rule
  Abstraction or Explicit Fragmentary Knowledge?}
\newblock \bibinfo{journal}{\emph{Journal of experimental psychology: General}}
  \bibinfo{volume}{119}, \bibinfo{number}{3} (\bibinfo{year}{1990}),
  \bibinfo{pages}{264}.
\newblock


\bibitem[Petersson et~al\mbox{.}(2012)]%
        {petersson2012artificial}
\bibfield{author}{\bibinfo{person}{Karl-Magnus Petersson},
  \bibinfo{person}{Vasiliki Folia}, {and} \bibinfo{person}{Peter Hagoort}.}
  \bibinfo{year}{2012}\natexlab{}.
\newblock \showarticletitle{{What Artificial Grammar Learning reveals about the
  Neurobiology of Syntax}}.
\newblock \bibinfo{journal}{\emph{Brain and language}} \bibinfo{volume}{120},
  \bibinfo{number}{2} (\bibinfo{year}{2012}), \bibinfo{pages}{83--95}.
\newblock


\bibitem[Pienemann et~al\mbox{.}(1988)]%
        {pienemann1988constructing}
\bibfield{author}{\bibinfo{person}{Manfred Pienemann}, \bibinfo{person}{Malcolm
  Johnston}, {and} \bibinfo{person}{Geoff Brindley}.}
  \bibinfo{year}{1988}\natexlab{}.
\newblock \showarticletitle{{Constructing an Acquisition-Based Procedure for
  Second Language Assessment}}.
\newblock \bibinfo{journal}{\emph{Studies in second language acquisition}}
  \bibinfo{volume}{10}, \bibinfo{number}{2} (\bibinfo{year}{1988}),
  \bibinfo{pages}{217--243}.
\newblock


\bibitem[Prat et~al\mbox{.}(2020)]%
        {prat2020relating}
\bibfield{author}{\bibinfo{person}{Chantel~S Prat}, \bibinfo{person}{Tara~M
  Madhyastha}, \bibinfo{person}{Malayka~J Mottarella}, {and}
  \bibinfo{person}{Chu-Hsuan Kuo}.} \bibinfo{year}{2020}\natexlab{}.
\newblock \showarticletitle{Relating natural language aptitude to individual
  differences in learning programming languages}.
\newblock \bibinfo{journal}{\emph{Scientific reports}} \bibinfo{volume}{10},
  \bibinfo{number}{1} (\bibinfo{year}{2020}), \bibinfo{pages}{3817}.
\newblock


\bibitem[Reber(1967)]%
        {reber1967implicit}
\bibfield{author}{\bibinfo{person}{Arthur Reber}.}
  \bibinfo{year}{1967}\natexlab{}.
\newblock \bibinfo{title}{{Implicit Learning of Artificial Grammars}}.
\newblock , \bibinfo{numpages}{855--863}~pages.
\newblock


\bibitem[Reber(1989)]%
        {reber1989implicit}
\bibfield{author}{\bibinfo{person}{Arthur Reber}.}
  \bibinfo{year}{1989}\natexlab{}.
\newblock \showarticletitle{{Implicit Learning and Tacit Knowledge}}.
\newblock \bibinfo{journal}{\emph{Journal of Experimental Psychology: General}}
  \bibinfo{volume}{118}, \bibinfo{number}{3} (\bibinfo{year}{1989}),
  \bibinfo{pages}{219}.
\newblock


\bibitem[Schachter(1983)]%
        {schachter1983new}
\bibfield{author}{\bibinfo{person}{Jacquelyn Schachter}.}
  \bibinfo{year}{1983}\natexlab{}.
\newblock \showarticletitle{{A New Account of Language Transfer}}.
\newblock \bibinfo{journal}{\emph{Language transfer in language learning}}
  \bibinfo{volume}{2} (\bibinfo{year}{1983}), \bibinfo{pages}{98--111}.
\newblock


\bibitem[Selinker(1969)]%
        {selinker1969language}
\bibfield{author}{\bibinfo{person}{Larry Selinker}.}
  \bibinfo{year}{1969}\natexlab{}.
\newblock \showarticletitle{{Language Transfer}}.
\newblock \bibinfo{journal}{\emph{General linguistics}} \bibinfo{volume}{9},
  \bibinfo{number}{2} (\bibinfo{year}{1969}), \bibinfo{pages}{67}.
\newblock


\bibitem[Shrestha et~al\mbox{.}(2018)]%
        {shrestha2018s}
\bibfield{author}{\bibinfo{person}{Nischal Shrestha}, \bibinfo{person}{Titus
  Barik}, {and} \bibinfo{person}{Chris Parnin}.}
  \bibinfo{year}{2018}\natexlab{}.
\newblock \showarticletitle{{It's like Python but: Towards Supporting Transfer
  of Programming Language Knowledge}}. In \bibinfo{booktitle}{\emph{Symposium
  on Visual Languages and Human-Centric Computing (VL/HCC)}}. IEEE,
  \bibinfo{pages}{177--185}.
\newblock


\bibitem[Shrestha et~al\mbox{.}(2020)]%
        {shrestha2020}
\bibfield{author}{\bibinfo{person}{Nischal Shrestha}, \bibinfo{person}{Colton
  Botta}, \bibinfo{person}{Titus Barik}, {and} \bibinfo{person}{Chris Parnin}.}
  \bibinfo{year}{2020}\natexlab{}.
\newblock \showarticletitle{{Here We Go Again: Why Is It Difficult for
  Developers to Learn Another Programming Language?}}. In
  \bibinfo{booktitle}{\emph{ICSE}}. \bibinfo{publisher}{IEEE},
  \bibinfo{pages}{691--701}.
\newblock


\bibitem[Siegmund et~al\mbox{.}(2014a)]%
        {siegmund2014understanding}
\bibfield{author}{\bibinfo{person}{Janet Siegmund}, \bibinfo{person}{Christian
  K{\"a}stner}, \bibinfo{person}{Sven Apel}, \bibinfo{person}{Chris Parnin},
  \bibinfo{person}{Anja Bethmann}, \bibinfo{person}{Thomas Leich},
  \bibinfo{person}{Gunter Saake}, {and} \bibinfo{person}{Andr{\'e} Brechmann}.}
  \bibinfo{year}{2014}\natexlab{a}.
\newblock \showarticletitle{{Understanding Understanding Source Code with
  Functional Magnetic Resonance Imaging}}. In \bibinfo{booktitle}{\emph{Proc.\
  Int'l Conf.\ Software Engineering (ICSE)}}. \bibinfo{publisher}{ACM},
  \bibinfo{pages}{378--389}.
\newblock


\bibitem[Siegmund et~al\mbox{.}(2014b)]%
        {siegmund2014measuring}
\bibfield{author}{\bibinfo{person}{Janet Siegmund}, \bibinfo{person}{Christian
  K{\"a}stner}, \bibinfo{person}{J{\"o}rg Liebig}, \bibinfo{person}{Sven Apel},
  {and} \bibinfo{person}{Stefan Hanenberg}.} \bibinfo{year}{2014}\natexlab{b}.
\newblock \showarticletitle{{Measuring and Modeling Programming Experience}}.
\newblock \bibinfo{journal}{\emph{Empirical Software Engineering}}
  \bibinfo{volume}{19}, \bibinfo{number}{5} (\bibinfo{year}{2014}),
  \bibinfo{pages}{1299--1334}.
\newblock


\bibitem[{Simon} et~al\mbox{.}(2019)]%
        {simon_pass_2019}
\bibfield{author}{\bibinfo{person}{{Simon}}, \bibinfo{person}{Andrew
  Luxton-Reilly}, \bibinfo{person}{Vangel Ajanovski}, \bibinfo{person}{Eric
  Fouh}, \bibinfo{person}{Christabel Gonsalvez}, \bibinfo{person}{Juho
  Leinonen}, \bibinfo{person}{Jack Parkinson}, \bibinfo{person}{Matthew Poole},
  {and} \bibinfo{person}{Neena Thota}.} \bibinfo{year}{2019}\natexlab{}.
\newblock \showarticletitle{{Pass {Rates} in {Introductory} {Programming} and
  in other {STEM} {Disciplines}}}. In \bibinfo{booktitle}{\emph{Proc. of the
  {Working} {Group} {Reports} on {Innovation} and {Technology} in {Computer}
  {Science} {Education}}}. \bibinfo{publisher}{ACM}, \bibinfo{address}{Aberdeen
  Scotland Uk}, \bibinfo{pages}{53--71}.
\newblock


\bibitem[Soloway et~al\mbox{.}(1982)]%
        {soloway_tapping_1982}
\bibfield{author}{\bibinfo{person}{Elliot Soloway}, \bibinfo{person}{Kate
  Ehrlich}, {and} \bibinfo{person}{Jeffrey Bonar}.}
  \bibinfo{year}{1982}\natexlab{}.
\newblock \showarticletitle{{Tapping into Tacit Programming Knowledge}}. In
  \bibinfo{booktitle}{\emph{Proceedings of the 1982 conference on {Human}
  factors in computing systems - {CHI} '82}}. \bibinfo{publisher}{ACM Press},
  \bibinfo{address}{Gaithersburg, Maryland, United States},
  \bibinfo{pages}{52--57}.
\newblock


\bibitem[Thorndike and Woodworth(1901)]%
        {Thorndike01}
\bibfield{author}{\bibinfo{person}{Edward Thorndike} {and}
  \bibinfo{person}{Robert Woodworth}.} \bibinfo{year}{1901}\natexlab{}.
\newblock \showarticletitle{{The Influence of Improvement in One Mental
  Function upon the Efficiency of Other Functions}}.
\newblock \bibinfo{journal}{\emph{Psychological Review}}  \bibinfo{volume}{8}
  (\bibinfo{year}{1901}), \bibinfo{pages}{247--261}.
\newblock
\newblock
\shownote{3}.


\bibitem[Tshukudu and Cutts(2020)]%
        {tshukudu2020understanding}
\bibfield{author}{\bibinfo{person}{Ethel Tshukudu} {and}
  \bibinfo{person}{Quintin Cutts}.} \bibinfo{year}{2020}\natexlab{}.
\newblock \showarticletitle{{Understanding Conceptual Transfer for Students
  Learning New Programming Languages}}. In \bibinfo{booktitle}{\emph{Proc.
  Conf. Int'l Computing Education Research}}. \bibinfo{publisher}{ACM},
  \bibinfo{pages}{227--237}.
\newblock


\bibitem[Utting et~al\mbox{.}(2013)]%
        {utting_fresh_2013}
\bibfield{author}{\bibinfo{person}{Ian Utting},
  \bibinfo{person}{Allison~Elliott Tew}, \bibinfo{person}{Mike McCracken},
  \bibinfo{person}{Lynda Thomas}, \bibinfo{person}{Dennis Bouvier},
  \bibinfo{person}{Roger Frye}, \bibinfo{person}{James Paterson},
  \bibinfo{person}{Michael Caspersen}, \bibinfo{person}{Yifat Ben-David
  Kolikant}, \bibinfo{person}{Juha Sorva}, {and} \bibinfo{person}{Tadeusz
  Wilusz}.} \bibinfo{year}{2013}\natexlab{}.
\newblock \showarticletitle{{A Fresh Look at Novice Programmers' Performance
  and Their Teachers' Expectations}}. In \bibinfo{booktitle}{\emph{{Proc. of
  the {ITiCSE} Working Group Reports Conference on {Innovation} and Technology
  in Computer Science Education-Working Group Reports}}}
  \emph{(\bibinfo{series}{{ITiCSE} -{WGR} '13})}.
  \bibinfo{publisher}{Association for Computing Machinery},
  \bibinfo{address}{New York, NY, USA}, \bibinfo{pages}{15--32}.
\newblock


\bibitem[Watson and Li(2014)]%
        {watson_failure_2014}
\bibfield{author}{\bibinfo{person}{Christopher Watson} {and}
  \bibinfo{person}{Frederick Li}.} \bibinfo{year}{2014}\natexlab{}.
\newblock \showarticletitle{{Failure Rates in Introductory Programming
  Revisited}}. In \bibinfo{booktitle}{\emph{Proc. Conf. on {Innovation} \&
  Technology in Computer Science Education}} \emph{(\bibinfo{series}{{ITiCSE}
  '14})}. \bibinfo{publisher}{ACM}, \bibinfo{address}{New York, NY, USA},
  \bibinfo{pages}{39--44}.
\newblock


\bibitem[Xie et~al\mbox{.}(2019)]%
        {xie_theory_2019}
\bibfield{author}{\bibinfo{person}{Benjamin Xie}, \bibinfo{person}{Dastyni
  Loksa}, \bibinfo{person}{Greg Nelson}, \bibinfo{person}{Matthew Davidson},
  \bibinfo{person}{Dongsheng Dong}, \bibinfo{person}{Harrison Kwik},
  \bibinfo{person}{Alex~Hui Tan}, \bibinfo{person}{Leanne Hwa},
  \bibinfo{person}{Min Li}, {and} \bibinfo{person}{Amy Ko}.}
  \bibinfo{year}{2019}\natexlab{}.
\newblock \showarticletitle{{A Theory of Instruction for Introductory
  Programming Skills}}.
\newblock \bibinfo{journal}{\emph{Computer Science Education}}
  \bibinfo{volume}{29}, \bibinfo{number}{2-3} (\bibinfo{year}{2019}),
  \bibinfo{pages}{205--253}.
\newblock


\end{thebibliography}

\end{document}